\documentclass[reprint,amsmath,amssymb,aip,]{revtex4-1}
\pdfoutput=1
\usepackage[latin1]{inputenc}
\usepackage[english]{babel}
\usepackage{amsmath}
\usepackage{amssymb}
\usepackage[pdftex]{graphicx}
\usepackage{dcolumn}
\usepackage{bm}
\usepackage{hyperref}

\begin{document}
\preprint{APS/123-QED}
\title{A complete laboratory for transport studies of electron-hole interactions in GaAs/AlGaAs systems}
\author{Ugo Siciliani de Cumis} 
\author{Joanna Waldie} 
\author{Andrew F. Croxall} 
\author{Deepyanti Taneja} 
\affiliation{Cavendish Laboratory, University of Cambridge, J.J. Thomson Avenue, Cambridge CB3 0HE, United Kingdom}

\author{Justin Llandro}
\altaffiliation[Present address: ]{Laboratory for Nanoelectronics and Spintronics, Research Institute of Electrical Communication, Tohoku University, Sendai 980-8577, Japan}
\affiliation{Cavendish Laboratory, University of Cambridge, J.J. Thomson Avenue, Cambridge CB3 0HE, United Kingdom}

\author{Ian Farrer}
\altaffiliation[Present address: ]{Department of Electronic and Electrical Engineering, University of Sheffield, 3 Solly Street, Sheffield S1 4DE, United Kingdom}
\affiliation{Cavendish Laboratory, University of Cambridge, J.J. Thomson Avenue, Cambridge CB3 0HE, United Kingdom}

\author{Harvey E. Beere} 
\author{David A. Ritchie}
\affiliation{Cavendish Laboratory, University of Cambridge, J.J. Thomson Avenue, Cambridge CB3 0HE, United Kingdom}

\date{\today}
\begin{abstract}
We present GaAs/AlGaAs double quantum well devices that can operate as both electron-hole (e-h) and hole-hole (h-h) bilayers, with separating barriers as narrow as 5 nm or 7.5 nm. With such narrow barriers, in the h-h configuration we observe signs of magnetic-field-induced exciton condensation in the quantum Hall bilayer regime. In the same devices we can study the zero-magnetic-field e-h and h-h bilayer states using Coulomb drag. Very strong e-h Coulomb drag resistivity (up to $10\%$ of the single layer resistivity) is observed at liquid helium temperatures, but no definite signs of exciton condensation are seen in this case. Self-consistent calculations of the electron and hole wavefunctions show this might be because the average interlayer separation is larger in the e-h case than the h-h case.
\end{abstract}
\maketitle
Systems of electrons and holes, which are confined in two different layers (bilayers), have been intensively studied in recent decades due to the possibility of formation of coherent phases of indirect excitons (i.e. whose fermionic components are spatially separated) \cite{lozovik1976new, shevchenko1976}. 
The attractive interaction between particles in different layers might lead to non-Fermi-liquid phases when the interlayer separation ($d$) becomes comparable with the mean intra-layer particle separation ($l$). Many different phases have been anticipated for such bilayer systems, from a condensate of indirect excitons with superfluid properties \cite{kulik1977excitonic, littlewood1996possibilities, balatsky2004dipolar}, to other possibilities  induced by localisation effects like charge density waves (CDW) \cite{liu1996static} or Wigner crystal-like solid states in one \cite{de2002excitonic, schleede2012phase}  or both layers \cite{schleede2012phase, joglekar2006wigner}. 

Most experimental attempts to test the theoretical expectations have exploited GaAs/AlGaAs double quantum well (DQW) structures, following three different approaches. First, optically generated indirect excitons in GaAs/AlGaAs systems have been studied, with a particular focus on macroscopic coherent ring-shaped patterns which suggest the presence of an excitonic condensate \cite{butov2002macroscopically, snoke2002long, high2012condensation}. 

The second method is to induce an electron-hole (e-h) bilayer by doping and/or electrostatic gating \cite{sivan1992coupled}. This class of devices benefits from having independent ohmic contacts to the two layers, making it possible to use transport experiments to probe the state of the system. In Coulomb drag measurements \cite{pogrebinskii1977mutual, price1983hot}, an electrical current is passed through one layer (the drive current) and an open-loop drag voltage forms across the second layer, as a result of interlayer momentum-energy exchange: the ratio between the drag electric field and the drive current density is known as drag resistivity and represents a direct measurement of the interlayer scattering. Such experiments have been performed on devices based on GaAs \cite{sivan1992coupled, croxall2008anomalous, seamons2009coulomb, zheng2016switching} and graphene structures \cite{kim2011coulomb, ponomarenko2011tunable, gorbachev2012strong, kim2012coulomb, gamucci2014anomalous}. An anomalous increase of the drag at low temperature was reported in several cases, suggesting the approach of a non-Fermi liquid phase; however, some unanswered questions still exist about the nature of this effect.

A third method uses hole-hole (h-h) and electron-electron (e-e) bilayers in double quantum well systems in a strong perpendicular magnetic field $B$. In this case, a phase transition is induced in the system when the total Landau level filling factor $\nu_T=\nu_1+\nu_2=1$ and the layers are sufficiently dilute ($d/l_B \lesssim 1.8$, where $l_B=(\hbar / eB)^{1/2}$ is the magnetic length \cite{murphy1994many}). The two layers become highly correlated because the Fermi level lies in the middle of the lowest Landau level in each layer, making it possible to consider the layers as both made of electrons or of holes (quantum Hall bilayers, QHB). 
Hence, the correlated state is analogous to an exciton condensate \cite{kellogg2002observation, kellogg2004vanishing, eisenstein2004bose, tutuc2004counterflow, nandi2012exciton, eisenstein2014exciton}.

Here, we present Coulomb drag measurements of electrically generated bilayer devices in a GaAs/AlGaAs DQW system. The separating barrier between the layers ($7.5$~nm or $5$~nm) should give stronger e-h interactions than in previous GaAs/AlGaAs e-h bilayers, where the barrier thickness was $\geq 10$~nm \cite{croxall2008anomalous, zheng2016switching}.
These devices can operate as either e-h or h-h bilayers, allowing us to look for exciton condensation using both the second and third approaches above in the same device. While the e-h bilayers show very strong Coulomb drag (up to 10\% of the single layer resistivity at temperature 3 K), there is no clear sign of exciton formation.
In the QHB regime the device with the 7.5~nm barrier in the h-h configuration shows clear evidence of exciton pairing with the expected signs of a condensate phase \cite{kellogg2002observation, tutuc2004counterflow}.
Having demonstrated exciton condensation in the h-h quantum Hall bilayer regime, we offer some remarks about how the exciton regime might be reached in this type of device operated as an e-h bilayer at zero magnetic field.

Our device, similar to those in Refs.~\onlinecite{seamons2007undoped} and \onlinecite{zheng2016switching}, is based on a completely undoped GaAs/AlGaAs DQW structure grown by Molecular Beam Epitaxy (MBE) on a semi-insulating (100)-oriented GaAs substrate.   
The GaAs quantum wells have a width of 15~nm. 
The mole fraction of AlAs in AlGaAs is $33\%$ with the exception of the barrier (thickness 5~nm or 7.5~nm) between the two quantum wells, where the concentration is increased to $90\%$ minimise interlayer leakage currents (normally less than 5$\%$ of the probing current).
	
\begin{figure}[t!]
	\centering
	\includegraphics[width=0.45\textwidth]{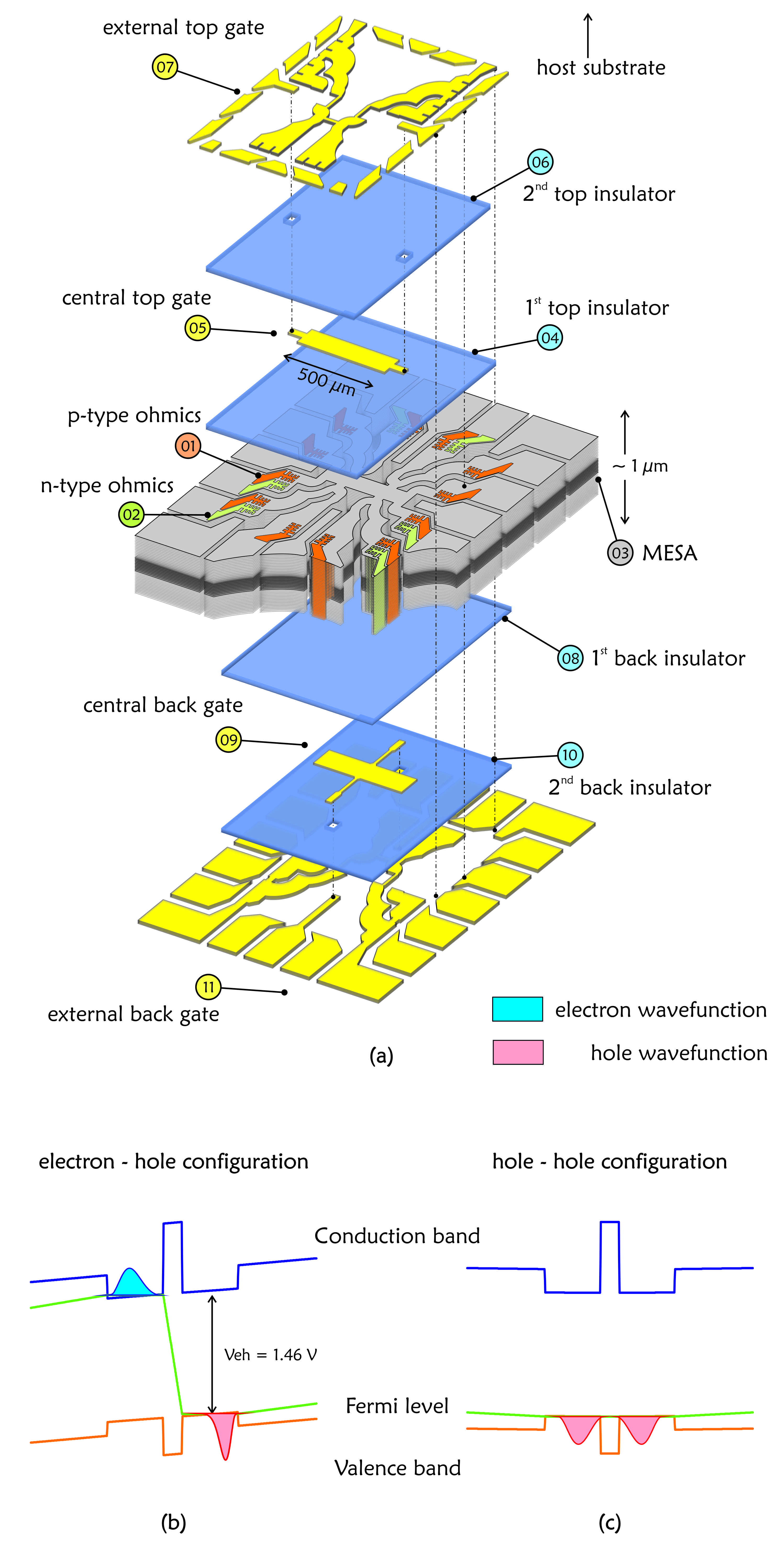}
	\caption{\textbf{(a)} Schematic of the layers of the device; the numbers in the circles refer to the order in which the layers are processed. \textbf{(b)} Energy band profile along the growth axis in the e-h and in the \textbf{(c)} h-h configuration. Diagrams in \textbf{(b)} and \textbf{(c)} were obtained by self-consistent calculations\cite{birner2011modeling} for a bilayer system with matched densities of $4\times 10^{10}$~cm$^{-2}$.}
	\label{sample_explosion}
\end{figure}
	
A schematic of a typical device is shown in Fig.~\textbf{\ref{sample_explosion}a}. Metal gates are required on both sides of the structure, to induce carriers in each quantum well, with independent control of the carrier density in each well. 
For an e-h bilayer, the two layers are kept at different chemical potentials, by applying a bias $V_{eh}$ between the two quantum wells. This moves the Fermi level from the valence band in one quantum well to the conduction band in the second quantum well (Fig.~\textbf{\ref{sample_explosion}b}) \cite{sivan1992coupled, seamons2007undoped, croxall2008anomalous}.
This bias is applied via annealed ohmic contacts ($p$-type AuBe and $n$-type AuGeNi). In this ambipolar design, one of the layers can host either electrons or holes (both $p$-type and $n$-type contacts are connected to this layer), making it possible to operate the device as either an e-h or a h-h bilayer (Fig.~\textbf{\ref{sample_explosion}c}) \cite{zheng2016switching}.
A 60-nm-thick Al$_2$O$_3$ dielectric layer, deposited by atomic layer deposition, is used to insulate the gates from the ohmic contacts.
In order to implement a double-side gated device, the GaAs substrate is completely removed\cite{weckwerth1996}, reducing the overall device thickness to less than 2~$\mu$m. 
The device is shaped in a double ``six-contacts" Hall bar (one for each layer). 
Typical mobilities in the reported devices are in the range $10^{5}$~-~$10^{6}$~cm$^{2}$ V$^{-1}$s$^{-1}$: the temperature dependence of the single layer resistivities was always metallic in the temperature range studied here. 

Transport experiments have been performed in a sorption-pumped $^3$He cryostat (minimum temperature 300~mK) and a $^3$He/$^4$He dilution refrigerator (down to 80~mK). 
Low-frequency ac (12~Hz) four-terminal Coulomb drag and magnetotransport measurements in a constant-current (5 or 10~nA) set-up were used to investigate the state of the system as a function of the temperature and the density of the layers.
The usual consistency checks for a linear relationship between drag voltage and drive current, scaling of the drag voltage with the Hall bar length-to-width ratio, and the effects of interlayer leakage \cite{gramila1991mutual, croxall2008anomalous} have been performed to exclude the contribution of spurious signals to the measured drag voltage (normally in the range of $\sim$nV).
Onsager's reciprocity theorem, applied to bilayer systems in the linear response regime, predicts that interchanging voltage and current probes in a Coulomb drag set-up at zero magnetic field should not affect the value of the drag resistivity \cite{casimir1945onsager}. In this text, the expressions electron drag and  hole drag correspond respectively to a current passing in the hole or in the electron layer. The reciprocity relation must be verified in an h-h bilayer as well: in this case those terms are replaced by back drag and top drag.

	\begin{figure}[t!]
		\centering
		\includegraphics[width=0.45\textwidth]{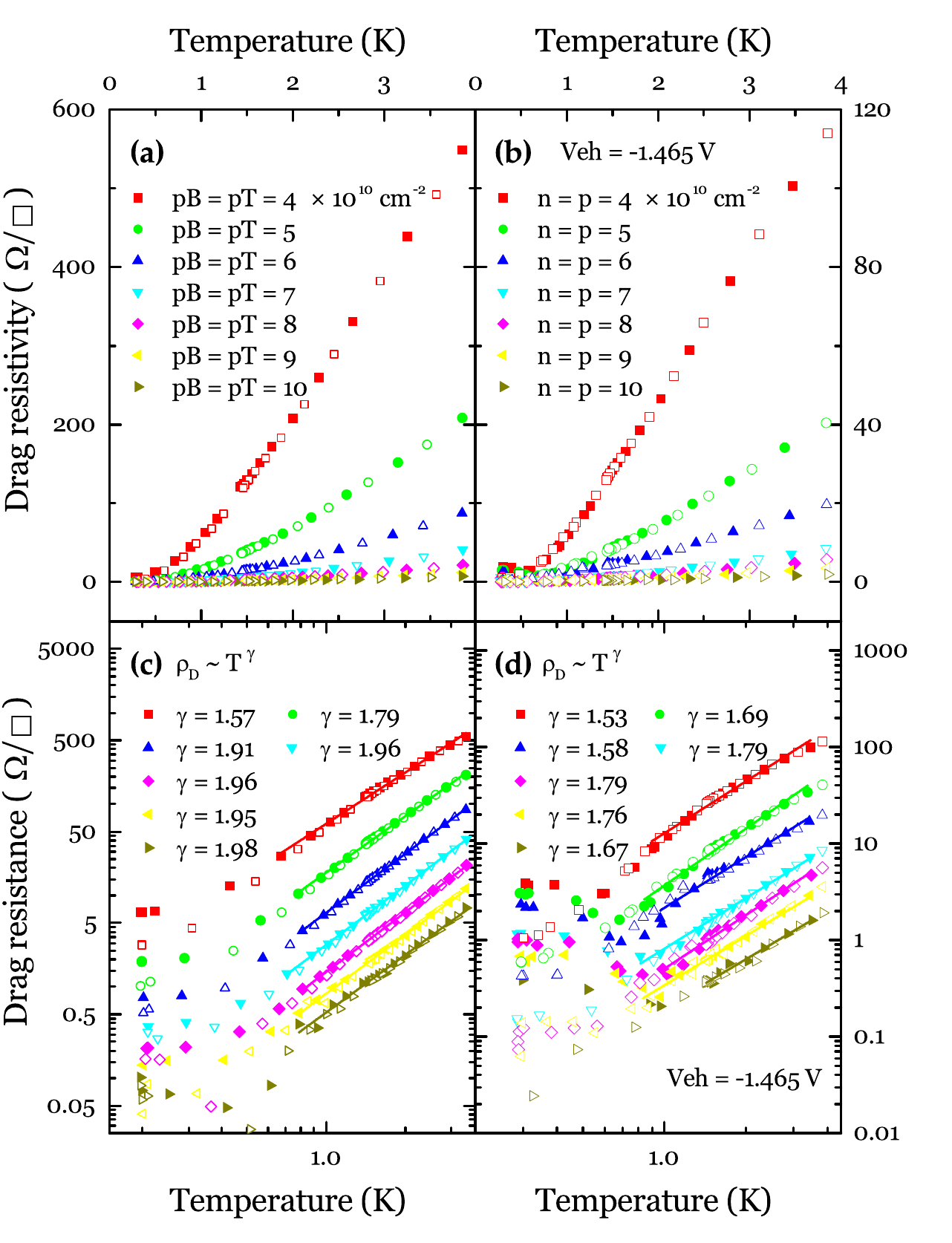}
		\caption{Drag resistivity as a function of temperature in a 7.5-nm-barrier device, in the \textbf{(a)} h-h and \textbf{(b)} e-h configurations. In the e-h (h-h) bilayer, full and empty symbols correspond respectively to hole(top) and electron(back) drag. The plots in \textbf{(c)} and \textbf{(d)} reproduce the same data as in \textbf{(a)} and \textbf{(b)} in a log-log scale to emphasise the high-density traces. The interlayer bias in the e-h configuration is $V_{eh}=-1.465$~V. Lines in \textbf{(c)} and \textbf{(d)} are non-linear fits of the function  $\rho_D \propto T^{\gamma}$.}
		\label{7.5_drag}
	\end{figure}
	\begin{figure}[h!]
		\centering
		\includegraphics[width=0.45\textwidth]{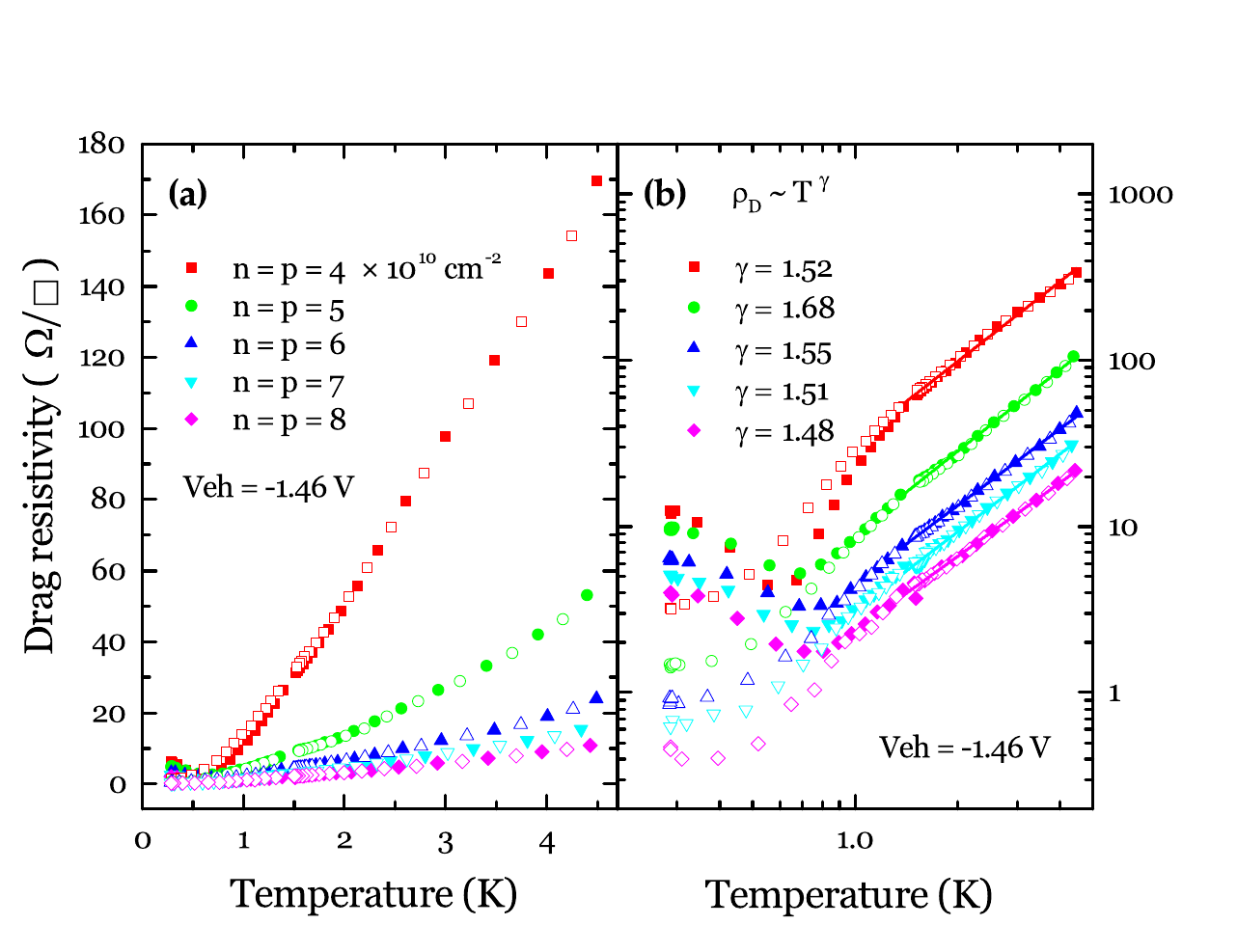}
		\caption{\textbf{(a)} Drag resistivity as a function of temperature in 5-nm-barrier device, in the e-h configuration. The interlayer bias is $V_{eh}=-1.46$~V; full and empty symbols correspond respectively to hole and electron drag. \textbf{(b)} Same data as in \textbf{(a)}, reproduced in a log-log scale to emphasise the high-density traces. Lines in \textbf{(b)} are non-linear fits of the function  $\rho_D \propto T^{\gamma}$.}
		\label{50_drag}
	\end{figure}

Figures \textbf{\ref{7.5_drag}a} and \textbf{\ref{7.5_drag}b} show drag resistivity ($\rho_D$) as a function of temperature ($T$) for respectively a h-h bilayer and an e-h bilayer activated in the same device during the same cool-down. The Al$_{0.9}$Ga$_{0.1}$As barrier width is 7.5~nm.  Consistently with what was reported by Zheng et al. \cite{zheng2016switching}, the h-h drag is bigger than the e-h drag at density (by approximately 5 times). This is partially due to the difference between the effective masses for electrons and holes, which causes the effective hole Bohr radius $a^*_B$ to be smaller than the electron one. Hence, the interaction parameter $r_s$ is higher for holes than for electrons (for a layer density of $n=4\times10^{10}$~cm$^{-2}$, $r_s=14$ for holes and $r_s=2.8$ for electrons). Here $r_s$ is the ratio between the intralayer Coulomb energy and the Fermi energy, $r_s= (a^*_B \cdot \sqrt{\pi n})^{-1}$. As a consequence, the screening of the hole layer is less effective than the electron one, causing the interlayer interaction to be stronger in the h-h case at equal temperature and densities. Moreover, the band-bending in the e-h bilayer tends to move the electron and the hole wavefunctions farther apart, whereas in the h-h bilayer this effect is less significant  (see Figs.~\textbf{\ref{sample_explosion}b} and \textbf{\ref{sample_explosion}c}). Self-consistent calculations for these devices with matched densities of $4\times10^{10}$~cm$^{-2}$ predict that the the peak-to-peak distance between the two wavefunctions is $\sim$22~nm for the h-h bilayer and $\sim$27~nm for the e-h case.

Figures \textbf{\ref{7.5_drag}c} and \textbf{\ref{7.5_drag}d} show the same data as Figs.~\textbf{\ref{7.5_drag}a} and \textbf{\ref{7.5_drag}b} on a log-log scale, in order to present more clearly the highest density and lowest temperature data. In the h-h bilayer the drag reciprocity is always verified. In the e-h case, the reciprocity is verified only for $T\gtrsim700$~mK. Below this temperature, an upturn in the hole drag signal starts to develop, whereas the electron drag decreases regularly to zero. Such a violation of Onsager's reciprocity theorem has been already observed in several other works \cite{croxall2008anomalous, seamons2009coulomb} and remains unexplained. In this non-reciprocal regime, although the drag voltage is still proportional to the drive current, some of the consistency tests fail (for instance, the drag resistance does not scale with the length-to-width ratio of the Hall bar). This suggests that the system is driven out of equilibrium, perhaps by electrical noise in the environment. A full comprehension of these effects has not been reached yet and further investigations are in progress.

Similar behaviour has been observed in devices with a 5-nm barrier (Fig.~\textbf{\ref{50_drag}a}).
In this sample, we have observed the largest reported drag resistivity in GaAs/AlGaAs e-h bilayers because of the extreme narrowness of the barrier. The drag resistivity at 3~K for layer densities of $4\times10^{10}$~cm$^{-2}$ is $\sim100~\Omega / \Box$, approximately $10 \%$ of the single-layer resistivity, making the drag mechanism significantly strong in these systems. 

It was not possible to achieve a stable h-h bilayer in any of the 5-nm-barrier samples tested, because of interlayer leakage currents. 
This could be because the absence of interlayer bias in the h-h configuration makes it possible for the hole wavefunctions to spread more uniformly in the quantum wells (see Figs.~\textbf{\ref{sample_explosion}b} and \textbf{\ref{sample_explosion}c}), increasing the wavefunction overlap across the barrier compared to the e-h case and, hence, the probability of interlayer tunnelling (in this device, the simulated peak-to-peak distance between the two wavefunctions is $\sim$19.5~nm for the h-h bilayer and $\sim$24.5~nm for the e-h case).

A non-linear fit of the relationship $\rho_D \propto T^{\gamma}$ has been used to quantify the temperature dependence of the drag in the regime of reciprocity ($1.2-4$~K).
For layer densities higher than  $6 \times 10^{10}$~cm$^{-2}$, the h-h drag approaches a $\rho_D \propto T^2$ regime, as expected for bilayers with sufficiently high density and large interlayer separation \cite{gramila1991mutual,jauho1993coulomb}. However, at the lowest density, the temperature dependence becomes slightly weaker, with $\gamma$ in the range 1.5-1.8. This is similar to previous results for low-density h-h bilayers, where a stronger than $T^2$-dependence was observed at low temperature, crossing over to a weaker than $T^2$-dependence at higher temperature \cite{pillarisetty2002frictional}. This behaviour can be explained within a Fermi liquid theory \cite{hwang2003frictional}.  In the e-h case, the parabolic regime is not even approached and $\gamma$ is always in the range 1.5-1.8 for both barrier widths. A more complete discussion about the temperature dependence of the drag resistivity will follow in a separate paper.
	
The h-h bilayer, in the sample with a 7.5-nm barrier, was explored in a strong perpendicular magnetic field. The quantum Hall bilayer regime is particularly sensitive to the interlayer separation and the densities: in previous attempts with a similar device, but with a 10-nm-barrier, the completely correlated state was not observed. On the other hand, this device with a narrower barrier shows signs of exciton condensation, as in previous experimental works on unipolar bilayers \cite{kellogg2002observation, tutuc2004counterflow, kellogg2003bilayer}. 

\begin{figure}[t!]
	\centering
	\includegraphics[width=0.45\textwidth]{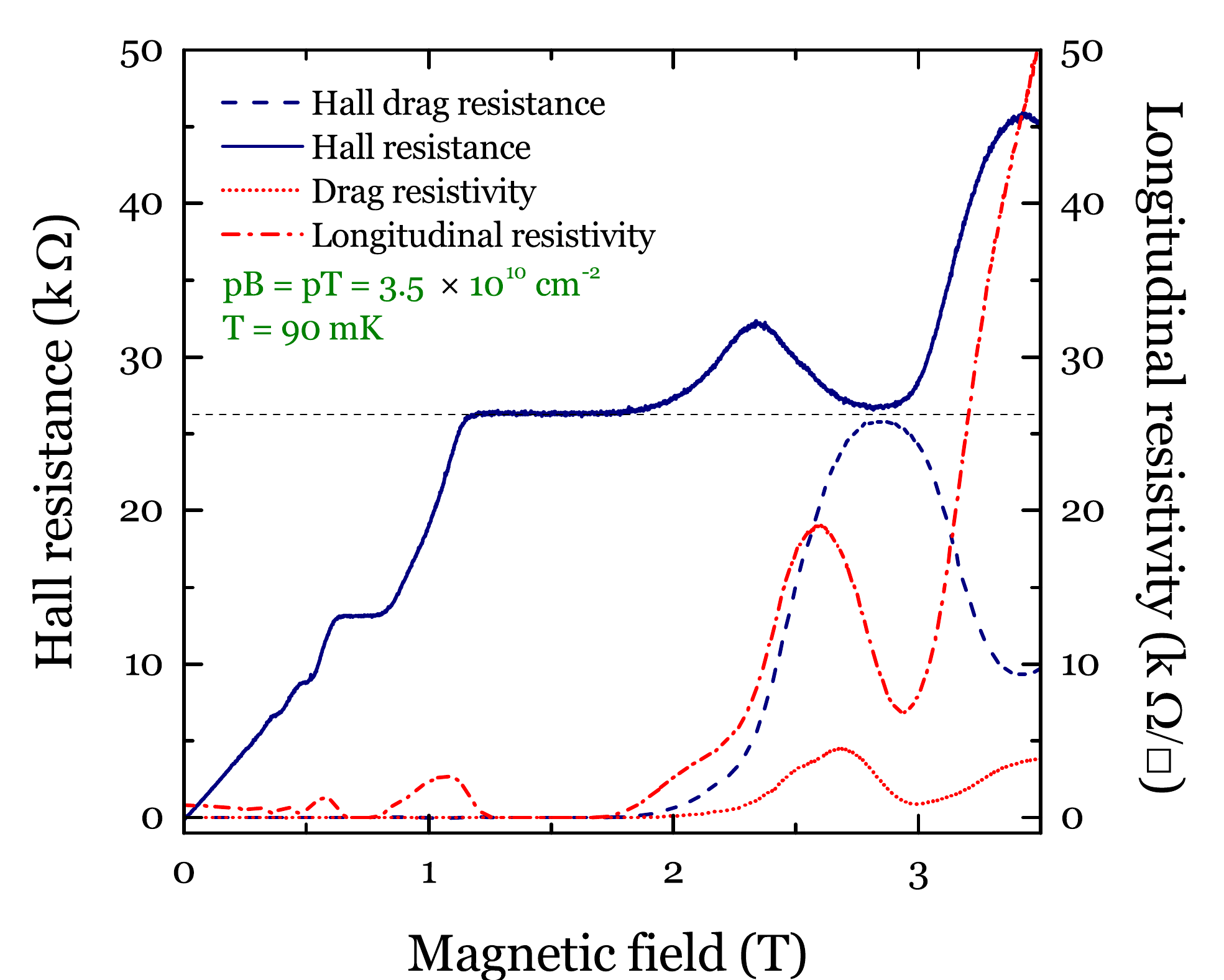}
	\caption{Hole-hole bilayer ($d/l_B\sim1.36$) in a perpendicular magnetic field. The current is passed in the top layer and Hall resistance and longitudinal resistivity are measured in both the drive and the drag layers. The $\nu_T=1$ state corresponds to $B\simeq2.9$~T.}
	\label{QHB_100mk}
\end{figure}

Fig.~\textbf{\ref{QHB_100mk}} reports the longitudinal resistivity and the Hall resistance in both the drive and drag layers as a function of $B$, with matched hole densities in the two layers $p_T=p_B=3.5 \times 10^{10}$~cm$^{-2}$ ($d/l_B=1.36$) and $T\approx90$~mK.
The drive current is 5~nA. 
For $B < 1.45$~T ($\nu_T > 2$) the traces corresponding to the drive layer follow the standard behaviour of a two-dimensional gas, exhibiting well defined quantum Hall plateaux and Shubnikov-de-Haas oscillations \cite{klitzing1980new} and the drag signals are small. 
However, when the system approaches $\nu_T=1$ ($B\simeq2.9$~T), the Hall resistances in both the drive and drag layers approach the same value on a plateau at $h/e^2$, the same level as at $\nu_T=2$ (which corresponds to $\nu=1$ in the single layers). In the same range, the longitudinal and the drag resistance begin to increase, before dropping to a minimum value at $\nu_T=1$. 
In the electron-hole picture, at $\nu_T=1$, each hole in one layer is bound to an \textit{electron-like} state in the other layer. The overlapping of the two Hall traces at $B\simeq2.9$~T is evidence of this effect.  
However, the longitudinal and the drag resistivity are expected to drop to zero in the quantum Hall bilayer state, corresponding to a dissipation-less motion of charge-neutral excitons. A dip in both signals is observable in Fig.~\textbf{\ref{QHB_100mk}}, although the effect is not as pronounced  as in the Hall resistance. 
This is probably due to the temperature in our experiments being higher than in previous studies \cite{tutuc2003role}.

These results demonstrate that excitonic correlations are visible in these devices in the h-h quantum Hall bilayer regime. However, no evidence of excitonic effects were observed in the e-h configuration at zero magnetic field. This may be because the mean interlayer separation is actually about 5~nm greater for an e-h bilayer than for an h-h bilayer generated in the same device at equal densities (see Figs.~\textbf{\ref{sample_explosion}b} and \textbf{\ref{sample_explosion}c}).  Reducing the barrier thickness even further, or make the quantum wells narrower, might give a chance of seeing excitons in the e-h configuration at zero field. 

When the e-h bilayer was tested at high magnetic field it was not possible to make reliable observations at $\nu_T=1$, due to disruptions in the normal functioning of the device. The bilayer is strongly affected by the capacitance between the layers and the interlayer bias. 
At high magnetic field the layer compressibilities and hence the interlayer capacitance are significantly modified, destabilising the e-h bilayer.
Measurements of the compressibility of the layers will help to better understand the effects of the quantum capacitance in this regime.
	
In conclusion, a set of ambipolar bilayer devices has been reported that can be operated as both an e-h and a h-h bilayer, with Al$_{0.9}$Ga$_{0.1}$As interlayer barriers of thickness 5~nm and 7.5~nm. 
The e-h drag resistivity at the lowest densities is approximately $10\%$ of the single layer resistivity, confirming that the system is approaching a regime of high e-h correlations. 
The devices with a barrier of 7.5~nm in the h-h configuration exhibited evidence of exciton condensation in the QHB regime at $\nu_T = 1$. Achieving lower densities than $4\times10^{10}$~cm$^{-2}$ would probably increase the chance of observing excitonic effects in the e-h configuration even at zero perpendicular magnetic field. 
\begin{acknowledgments}
This work was funded by EPSRC EP/H017720/1 and EP/J003417/1 and  European Union Grant INDEX 289968. A.F.C. acknowledges funding from Trinity College, Cambridge, and D.T. from St. Catherine's College, Cambridge. I.F. acknowledges funding from Toshiba Research Europe Limited.
\end{acknowledgments}
%
	
%

\end{document}